\documentclass[aps, prb, twocolumn, showpacs] {revtex4}
\usepackage{epsfig,amsfonts,amssymb,amsmath}

\begin{document}

%%%%%%%%%%%%%%%%%%%%%%%%%%%%%%%%%%%%%%%%%%%%%%%%%%%%%%%%%%%%%%%%%
% Defintions   
%%%%%%%%%%%%%%%%%%%%%%%%%%%%%%%%%%%%%%%%%%%%%%%%%%%%%%%%%%%%%%%%%
\renewcommand{\Re}{\operatorname{Re}}
\renewcommand{\Im}{\operatorname{Im}}
\newcommand{\Tr}{\operatorname{Tr}}
\newcommand{\sign}{\operatorname{sign}}
\newcommand{\dd}{\text{d}}
\newcommand{\q}{\boldsymbol q}
\newcommand{\p}{\boldsymbol p}
\newcommand{\rr}{\boldsymbol r}
\newcommand{\pp}{p_v}
\newcommand{\vv}{\boldsymbol v}
\newcommand{\I}{{\rm i}}
\newcommand{\pphi}{\boldsymbol \phi}
\newcommand{\ds}{\displaystyle}
\newcommand{\be}{\begin{equation}}
\newcommand{\ee}{\end{equation}}
\newcommand{\bea}{\begin{eqnarray}}
\newcommand{\eea}{\end{eqnarray}}
\newcommand{\Acl}{{\cal A}}
\newcommand{\Rcl}{{\cal R}}
\newcommand{\Tcl}{{\cal T}}
\newcommand{\Tmin}{{T_{\rm min}}}
\newcommand{\Toff}{{\langle \delta T \rangle_{\rm off} }}
\newcommand{\Roff}{{\langle \delta R \rangle_{\rm off} }}
\newcommand{\RoffI}{{\langle \delta R_I \rangle_{\rm off} }}
\newcommand{\RoffII}{{\langle \delta R_{II} \rangle_{\rm off} }}
\newcommand{\dg}{{\langle \delta g \rangle_{\rm off} }}
\newcommand{\rd}{{\rm d}}
\newcommand{\br}{{\bf r}}
\newcommand{\la}{\langle}
\newcommand{\ra}{\rangle}
\newcommand{\ua}{\uparrow}
\newcommand{\da}{\downarrow}
%%%%%%%%%%%%%%%%%%%%%%%%%%%%%%%%%%%%%%%%%%%%%%%%%%%%%%%%%%%%%%%%%%%

\title{Entanglement production in chaotic quantum dots subject to spin-orbit coupling} 

\author{Diego Frustaglia$^1$, Simone Montangero$^1$, and Rosario Fazio$^{1,2}$}  
\affiliation{$^1$NEST-CNR-INFM \& Scuola Normale Superiore, I-56126 Pisa, Italy\\
$^2$International School for Advanced Studies (SISSA), I-34014, Trieste, Italy}

\date{\today}

\begin{abstract}

We study numerically the production of orbital and spin entangled states in chaotic quantum dots
for non-interacting electrons. 
The introduction of spin-orbit coupling permit us to identify 
signatures of time-reversal symmetry correlations in the entanglement production 
previously unnoticed, 
resembling weak-(anti)localization quantum corrections to the conductance. 
We find the entanglement to be strongly dependent on  
spin-orbit coupling, showing universal features for broken time-reversal and 
spin-rotation symmetries. 

\end{abstract} 


% 72.25.Rb Spin relaxation and scattering
% 71.70.Ej Spin-orbit coupling, Zeeman and Stark splitting, Jahn-Teller effect
% 05.45.Pq Numerical simulations of chaotic systems
% 03.67.Mn Entanglement production, characterization, and manipulation
% 72.10.-d Theory of electronic transport; scattering mechanisms
% 73.23.-b Electronic transport in mesoscopic systems

\maketitle

%%%%%%%%%%%%%%%%%%%%%%%%%%%%%%%%%%%%%%%%%%%%%%%%%%%%%%%%%%%%%%%%%%%%%
% BODY OF PAPER
%%%%%%%%%%%%%%%%%%%%%%%%%%%%%%%%%%%%%%%%%%%%%%%%%%%%%%%%%%%%%%%%%%%%%

%====================================================================
\section{Introduction}
%====================================================================

The existence of \emph{entangled} many-particle quantum states subject to 
nonclassical correlations is widely recognized 
as a fundamental resource for quantum information processing. 
Two quantum systems $A$ and $B$ are said to be entangled if they are 
\emph{not separable}, namely, if the common (pure) state $| \Psi_{AB} \rangle$ 
\emph{can not} be written as the product of individual states $|\Psi_A \rangle$
and $|\Psi_B \rangle$
(i.e., $|\Psi_{AB} \rangle \neq |\Psi_A \rangle |\Psi_B \rangle$). 
In such case the internal degrees of freedom of systems $A$ and $B$ are
quantum mechanically correlated, since any measurement performed on system 
$A$ would condition the results of a measurement on system $B$ beyond any
classical constraint.
In solid-state 
physics, several aspects related to the production, control and detection of 
entangled electronic states have been addressed (see Ref.~\onlinecite{beenakker05} 
for a recent review) and an extended literature
already exists.\cite{loss01,lesovik01,samuelsson03,prada04,
sauret04,recher02,bena01,bouchiat03,oliver02,saraga03,costa01,beenakker03-2,beenakker03,
samuelsson04,beenakker04,beenakker04-2,samuelsson04-3,saraga04,faoro04,signal05,egues05,
samuelsson05,bose02,lebedev05} 
Most of the existing proposals for the generation of electronic entanglement 
are based on the presence of some kind of interaction between the particles 
involved as, e.g., Coulomb interaction, (anti)ferromagnetism, superconducting pairing, etc.
In contrast to the original belief, it was recently recognized that interactions 
are indeed not necessary to 
produce entanglement. Non-interacting electrons, initially in a separable 
uncorrelated state, can evolve into an entangled state due to exchange correlations in a 
scattering process from an external 
potential.\cite{beenakker03,samuelsson04,beenakker03-2,samuelsson05,beenakker05,signal05,bose02,lebedev05} 
This applies to electronic transport in multiterminal mesoscopic quantum conductors.
In this approach, the efficiency of the entangler depends on the particular 
characteristics of the scatterer, which is described by its corresponding 
scattering matrix $S$. 
Among the several possibilities, systems of special interest are disorder-free 
chaotic quantum dots, also referred to as chaotic billards. These have the advantage of allowing 
for a statistical analysis
that can reveal universal properties of (chaotic) electronic entanglers. 
This case was recently addressed by Beenakker {\it et al.}\cite{beenakker03-2} by 
means of a random-matrix-theory (RMT) approach. 
They obtained a universal mean 
value for the degree of two-electron entanglement produced between spatially 
separated {\it orbital} channels, and remarkably found that it is not 
significatively affected by the breaking of time-reversal symmetry (TRS)- in contrast
to other (single-particle) transport characteristics as the conductance.
Later, Samuelsson, Sukhorukov, and B\"uttiker\cite{samuelsson05} reformulated the problem 
by formally including the spin, though in the absence of any spin-dependent interaction.  

In this paper we study the production of both orbital and spin two-particle entanglement 
for non-interacting electrons in multiterminal chaotic billards. We approach the problem 
numerically. 
By introducing the Rashba spin-orbit (SO) coupling\cite{bychkov84} together with 
a magnetic flux breaking TRS, we are able to identify the signatures of 
weak-localization (WL) and weak-antilocalization (WA) quantum corrections  
in the entanglement production. Such TRS-correlation effects appear as a constraint
that can limit/enhance the efficiency of the entangler.
We find that the production of spin as well as orbital entanglement is strongly 
affected by SO coupling on a scale corresponding to the pass from WL to WA. 
Additionally, we show that a finite residual entanglement survives after the 
breaking of both TRS and spin-rotation symmetry showing some universal 
characteristics.

The paper is organized as follows.
In Sec.~\ref{model-SP} we describe our numerical model and shortly review some 
relevant single-particle properties as the WL and WA quantum corrections to 
the conductance. 
In Sec.~\ref{ent-2P} we start discussing the interaction-free production of 
entanglement from a separable two-particle state (Sec.~\ref{ent-2P-a}). We introduce 
the concurrence as a measure of two-qubit entanglement in Sec.~\ref{ent-2P-b}.
In Sec.~\ref{ParCon} we discuss some features related to the accessible entanglement 
as a consequence of the local particle number conservation.
The results are presented in Sec.~\ref{results}, followed by a short 
summary of conclusions in Sec.~\ref{conclu}. 

%%%%%%%%%%%%%%%%%%%%%%%%%%%%%%%%%%%%%%%%%%%%%%%%%%%%%%%%%%%%%%%%%%%%%%%%%%%%%%%%%%%%%%%%%%
%                                       FIGURE
%%%%%%%%%%%%%%%%%%%%%%%%%%%%%%%%%%%%%%%%%%%%%%%%%%%%%%%%%%%%%%%%%%%%%%%%%%%%%%%%%%%%%%%%%%
\begin{figure}
%%[tbp]
\includegraphics[width=.35 \textwidth, angle=0]{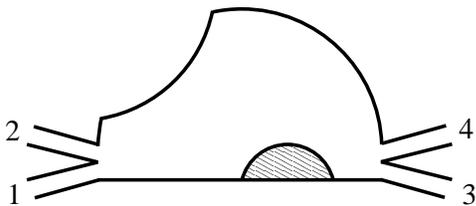}
\caption{Chaotic quantum dot entangler used in the numerical model. Leads connected to
electron reservoirs support one orbital channel plus two spin channels each.
}
\label{fig-1}
\end{figure}
%%%%%%%%%%%%%%%%%%%%%%%%%%%%%%%%%%%%%%%%%%%%%%%%%%%%%%%%%%%%%%%%%%%%%%%%%%%%%%%%%%%%%%%%%%

%====================================================================
\section{Model and single-particle properties}
\label{model-SP}
%====================================================================

We consider a two-dimensional chaotic dot 
connected to electron reservoirs at the left and right, as shown in 
Fig.~\ref{fig-1}. \cite{note3} 
Two single-orbital-channel leads are attached at each side of the dot, similar to 
what proposed in Ref.~\onlinecite{beenakker03-2}. 
In addition, each orbital mode can support two spin channels.
The single-particle Hamiltonian for electrons with charge $-e$ and effective mass $m^*$
reads
\be
H=\frac {\boldsymbol{\Pi}^2}{2 m^*}+
\frac {\alpha_{\rm R}}{\hbar} \left(\boldsymbol{\sigma} \times
\boldsymbol{\Pi}\right)_z+V({\bf r}),
\ee
where $\boldsymbol{\Pi}={\bf p} + (e/c) {\bf A}$, $\alpha_{\rm R}$ is the Rashba SO
coupling strength, $\boldsymbol{\sigma}$ is the vector of Pauli spin matrices, 
$z$ is the axis perpendicular to the dot's plane, and $V$ is a hard-wall confining 
potential describing the dot's contour.
The strength of the Rashba SO interaction can be given
in terms of the ratio $L_{\rm esc}/L_{\rm SO}$, where 
$L_{\rm esc}=\pi A/w$ is the classical escape length in an open chaotic billard 
of area $A$ with a total opening of width $w$,\cite{RS02} and 
$L_{\rm SO}=\pi \hbar^2 /\alpha_{\rm R} m^*$ 
is the spin-precession length due to SO coupling. 
A uniform magnetic field (generated by the vector potential ${\bf A}$) introduces a 
flux $\phi$ (measured in units of the flux quantum $\phi_0=h c/e$).
Applying a (small) bias voltage between reservoirs produces a coherent
electron current through the dot from left to right. 
For the calculation of the corresponding scattering amplitudes we implement a 
recursive Green's function technique based on a spin-dependent tight-binding 
model arising from a real-space discretization.\cite{note4} Additionally, we also 
perform some independent RMT simulations for comparison.

\begin{figure}
%%[tbp]
\includegraphics[width=.45 \textwidth, angle=0]{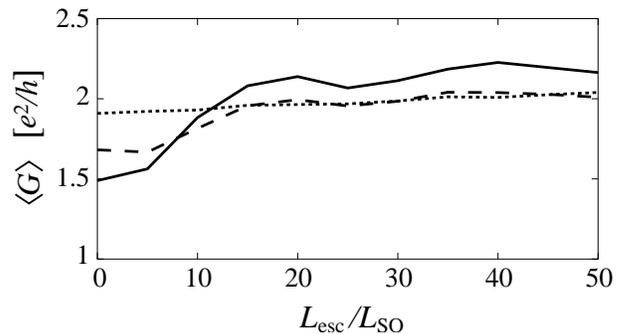}
\caption{Sample-averaged conductance $\langle G \rangle$ for the chaotic 
quantum dot of Fig.~\ref{fig-1} vs. SO coupling in the presence of a magnetic flux $\phi$. 
The curves correspond to $\phi/\phi_0=0$ (solid line), $\phi/\phi_0=1/2$ (dashed line), and 
$\phi/\phi_0=5$ (dotted line).
The results illustrate WL (WA) quantum corrections to the 
classical conductance ($2 e^2/h$ in this case) for
weak (large) SO coupling due to TRS ($\phi/\phi_0=0$).
These quantum correction disappear as TRS breaks ($\phi/\phi_0 \neq 0$).}
\label{fig-2}
\end{figure}

The single-particle transport properties are 
characterized by the Landauer-B\"uttiker linear conductance $G$.
For illustration, in Fig.~\ref{fig-2} we present some numerical results for the 
sample-average conductance $\langle G \rangle$ of the chaotic dot of Fig.~\ref{fig-1}
as a function of $L_{\rm esc}/L_{\rm SO}$ and $\phi$.
Coherent backscattering leads to a \emph{minimum} in $G$ (WL) at $\phi=0$ in the absence of 
SO coupling. For large SO coupling, $G$ presents a \emph{maximum} (WA) instead.  
The transition from WL to WA shows up around $L_{\rm esc}/L_{\rm SO} \approx 10$ 
(see also Refs.~\onlinecite{zumbuhl02,AF01,BCH02,zaitsev05}). For a large $\phi$, 
TRS is broken and $G$ is independent of the SO coupling 
stregth, remaining close to its classical value ($G_{\rm cl}=2e^2/h$ in our case). 
Such underlying single-particle physics is relevant for the understanding of the new 
two-particle effects reported in this paper.
The crossover from WL to WA also manifest, though differently, in the entanglement 
production as we see below.

%%%%%%%%%%%%%%%%%%%%%%%%%%%%%%%%%%%%%%%%%%%%%%%%%%%%%%%%%%%%%%%%%%%%%%%%%%%%%%%%%%%%%%%%%%
%                                       FIGURE
%%%%%%%%%%%%%%%%%%%%%%%%%%%%%%%%%%%%%%%%%%%%%%%%%%%%%%%%%%%%%%%%%%%%%%%%%%%%%%%%%%%%%%%%%%
%%%%%%%%%%%%%%%%%%%%%%%%%%%%%%%%%%%%%%%%%%%%%%%%%%%%%%%%%%%%%%%%%%%%%%%%%%%%%%%%%%%%%%%%%%

%====================================================================
\section{Entanglement production from separable two-particle states}
\label{ent-2P}
%====================================================================

%--------------------------------------------------------------------
\subsection{Incoming and outgoing two-particle states}
\label{ent-2P-a}
%--------------------------------------------------------------------

We consider a separable two-particle state incoming from an electron 
reservoir on the left of the dot of Fig.~\ref{fig-1}: 
\be
| \Psi_{\rm in} \ra = a_1^{s_1 \dag}  a_2^{s_2 \dag} | 0 \rangle,
\label{IN}
\ee 
where $a_i^{s_i \dag}$ creates an incoming electron in lead $i=1,2$ with spin $s_i=\ua,\da$,
and $| 0 \rangle$ is the Fermi sea at zero temperature.
Multiple scattering within the dot entangles the ougoing state. This is a coherent 
superposition of orbital and spin channels determined by the single-particle $S$-matrix.
It reads
\be
| \Psi_{\rm out} \rangle = \sum_{n, \sigma} \sum_{m, \tau} S_{n 1}^{\sigma s_1}
S_{m 2}^{\tau s_2}~b_n^{\sigma \dag}  b_m^{\tau \dag} | 0 \rangle,
\label{OUT}
\ee
where $S_{j i}^{s s_i}$ is the scattering amplitude from 
lead $i=1,2$ with spin $s_i$ to any lead $j=1,...,4$ with spin $s$.
The $b_j^{s \dag}$ creates an outgoing electron in lead $j$ with spin $s$,
satisfying the vector equation $b^\dag \cdot S= a^\dag$. 
The terms in (\ref{OUT}) with $n=m$, $\sigma=\tau$ vanish for the sake of Fermionic 
statistics. 

%--------------------------------------------------------------------
\subsection{Particle conservation}
\label{ParCon}
%--------------------------------------------------------------------

The $| \Psi_{\rm out} \ra$ of Eq.~(\ref{OUT}) can be split into three terms 
with different local particle number at the left ($n_{\rm L}$) and right 
($n_{\rm R}$) of the dot such that $n_{\rm L}+n_{\rm R}=2$, in the form
\bea
\nonumber
| \Psi_{\rm out} \ra &=& \sum_{n_{\rm L},n_{\rm R}} | n_{\rm L},n_{\rm R}\ra \\
&=&|2,0 \ra + |0,2 \ra + |1,1 \ra. 
\label{sOUT}
\eea
The accessible entanglement\cite{wiseman03} in $| \Psi_{\rm out} \ra$ is studied 
in Bell-like measurements by performing local operations that conserve the local particle 
number.\cite{beenakker05} This is of fundamental importance for electrons, preventing 
the local creation of states in a coherent superposition of different number of particles.
This means that local operations do not mix the three terms of Eq.~(\ref{sOUT})
and the entanglement can be studied in each of them \emph{independently}. 

%--------------------------------------------------------------------
\subsection{The concurrence as an entanglement measure}
\label{ent-2P-b}
%--------------------------------------------------------------------

We evaluate the amount of entanglement between pairs of two-level (sub)systems or 
\emph{qubits}, which in our case correspond to an electron leaving the quantum dot 
in one of two predetermined orbital/spin states. 
This requires a bipartition of the system by
choosing some pairs of outgoing channels of interest, 
tracing out the nonobserved degrees of freedom compatible with that choice.
The two-qubit entanglement is quantified by the \emph{concurrence} $0 \le C \le 1$, 
defined as\cite{wootters98}
\be
C(\rho) \equiv \max \{0,\lambda_1-\lambda_2-\lambda_3-\lambda_4\}.
\label{conc}
\ee
The $\lambda_i$s are the square roots of the eigenvalues (in decreasing order) 
of the matrix $\rho \tilde{\rho}$, where $\rho \in 4\times 4$ is a two-qubit density 
matrix and $\tilde{\rho} \equiv (\sigma_y \otimes \sigma_y) \rho^* (\sigma_y \otimes \sigma_y)$,
with $\sigma_y$ the second Pauli matrix. Separable unentangled states have $C=0$, while
$C=1$ correspond to maximally entangled (Bell) states.
States with $0 < C < 1$ are non-separable, partly entangled states.
A $C \neq 0$ is a necessary and sufficient condition for affirming that the 
two involved qubits are non-classically correlated due to entanglement.
For electrons in the quantum dot of Fig.~\ref{fig-1}, where both orbital and spin degrees 
of freedom are involved, the two-particle density matrix $\rho$ is generally larger than 
$4 \times 4$, i.e., it does not correspond to a two-qubit system. Still, a study of 
the, e.g., orbital qubits can be performed by defining a reduced density matrix (RDM) 
${\rm Tr}_s (\rho) \in 4 \times 4$, where the trace is taken over the spin 
degree of freedom $s$; and vice versa (see Sec.~\ref{results}). 
For a chaotic dot, the outgoing wave function (\ref{OUT}) is a function of a 
\emph{random} scattering matrix $S$, and so its corresponding density matrix. 
The entanglement contained in (\ref{OUT}) depends on the particular $S$, which is
sample dependent. Hence, we characterize the production of entanglement in by 
calculating the sample-averaged $\la C \ra$ and its fluctuations 
${\rm var}(C) \equiv \la C^2 \ra-\la C \ra^2$. 
The concurrence can be related to zero-frequency current-noise measurements 
(i.e., without time-resolved detection) in mesoscopic conductors, as recently shown 
in Ref.~\onlinecite{beenakker05} and \onlinecite{beenakker03-2}.
Alternatively, the presence of entanglement could be determined by other means as,
e.g., beam-splitter current correlations giving a lower bound for entanglement\cite{BL03-GFTF06}
or implementing some entanglement witness.\cite{witness}
Here, we calculate $C$ from its definition (\ref{conc}) for several
bipartitions of the outgoing state (\ref{OUT}) independently of any 
particular detection scheme.

%====================================================================
\section{Results}
\label{results}
%====================================================================

%%%%%%%%%%%%%%%%%%%%%%%%%%%%%%%%%%%%%%%%%%%%%%%%%%%%%%%%%%%%%%%%%%%%%%%%%%%%%%%%%%%%%%%%%%
%                                       FIGURE
%%%%%%%%%%%%%%%%%%%%%%%%%%%%%%%%%%%%%%%%%%%%%%%%%%%%%%%%%%%%%%%%%%%%%%%%%%%%%%%%%%%%%%%%%%
\begin{figure}
%%[tbp]
\includegraphics[width=.4 \textwidth, angle=0]{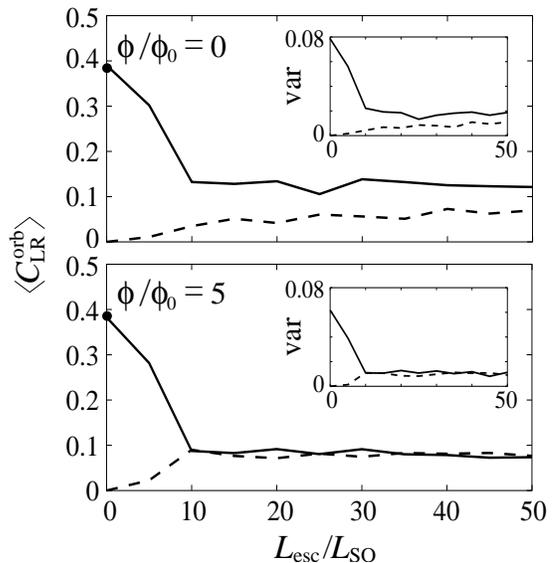}
\caption{Left-right orbital entanglement: $\la C_{\rm LR}^{\rm orb} \ra$ vs SO coupling  
with TRS preserved (upper panel) and broken (lower panel). Solid (dashed) lines correrspond
to (anti)parallel incoming spins. Insets depict the fluctuations ${\rm var}(C)$, respectivelly.
Full dots: RMT results from Ref.~\onlinecite{beenakker03-2}.
}
\label{fig-3}
\end{figure}
%%%%%%%%%%%%%%%%%%%%%%%%%%%%%%%%%%%%%%%%%%%%%%%%%%%%%%%%%%%%%%%%%%%%%%%%%%%%%%%%%%%%%%%%%%

%%%%%%%%%%%%%%%%%%%%%%%%%%%%%%%%%%%%%%%%%%%%%%%%%%%%%%%%%%%%%%%%%%%%%%%%%%%%%%%%%%%%%%%%%%
%                                       FIGURE
%%%%%%%%%%%%%%%%%%%%%%%%%%%%%%%%%%%%%%%%%%%%%%%%%%%%%%%%%%%%%%%%%%%%%%%%%%%%%%%%%%%%%%%%%%
\begin{figure}
%%[tbp]
\includegraphics[width=.4 \textwidth, angle=0]{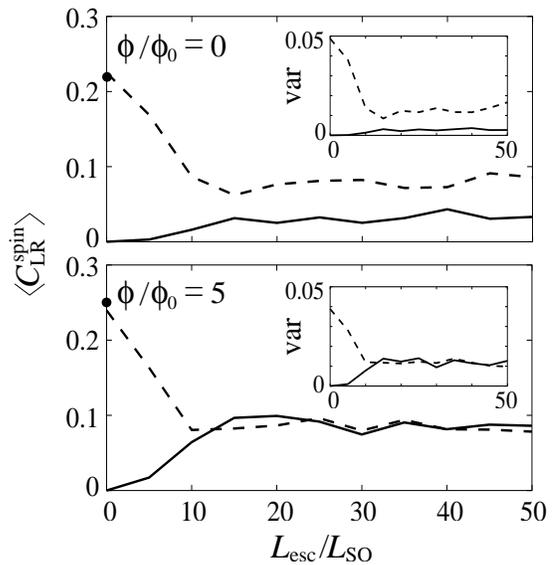}
\caption{Left-right spin entanglement: $\la C_{\rm LR}^{\rm spin} \ra$ vs. SO coupling  
with TRS preserved (upper panel) and broken (lower panel). Solid (dashed) lines correrspond
to (anti)parallel incoming spins. Fluctuations ${\rm var}(C)$ shown in insets.
Full dots: RMT results.
}
\label{fig-4}
\end{figure}
%%%%%%%%%%%%%%%%%%%%%%%%%%%%%%%%%%%%%%%%%%%%%%%%%%%%%%%%%%%%%%%%%%%%%%%%%%%%%%%%%%%%%%%%%%

\begin{figure}
%%[tbp]
\includegraphics[width=.4 \textwidth, angle=0]{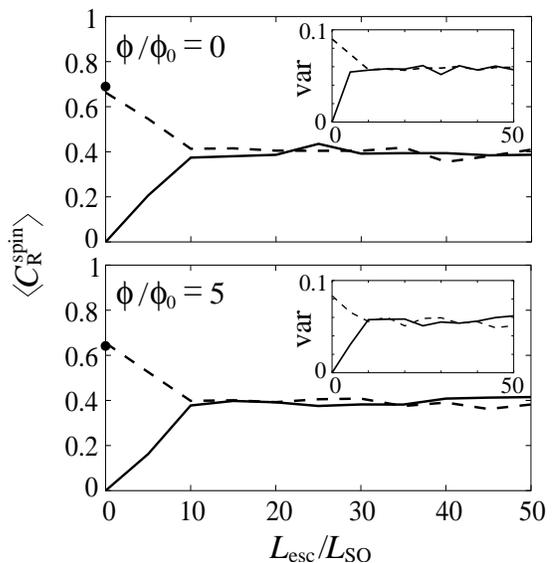}
\caption{Transmitted spin entanglement: $\la C_{\rm R}^{\rm spin} \ra$ vs. 
SO coupling with TRS preserved (upper panel) and broken (lower panel). 
Solid (dashed) lines correrspond to (anti)parallel incoming spins. Insets show the fluctuations
${\rm var}(C)$. Full dots: RMT results.
}
\label{fig-5}
\end{figure}

We investigate first the entanglement produced between outgoing left and right 
channels. To this aim we project $| \Psi_{\rm out} \ra$ onto the subspace 
containing one single excitation at each side of the dot (i.e. $|1,1 \ra$). 
We obtain 
\be
| \Psi_{\rm LR} \ra \equiv 
\sum_{p,\alpha} \sum_{q,\beta} ( S_{p 1}^{\alpha s_1} S_{q 2}^{\beta s_2} - S_{q 1}^{\beta s_1} S_{p 2}^{\alpha s_2} )~b_p^{\alpha \dag}  b_q^{\beta \dag} |0 \rangle,
\label{outLR}
\ee 
where the indices $p=1,2;\alpha=\ua,\da$ and $q=3,4;\beta=\ua,\da$ stand for left and right 
outgoing channels, respectively. 
The density matrix of the state (\ref{outLR}) reads 
$\rho_{\rm LR} = |\Psi_{\rm LR} \ra \la \Psi_{\rm LR}|/\la \Psi_{\rm LR}|\Psi_{\rm LR} \ra \in 8 \times 8$,
which by construction accounts for both orbital and spin degrees of freedom. The degree of 
{\it orbital entanglement} contained in (\ref{outLR}) can be extracted from $\rho_{\rm LR}$ 
by tracing out the spin degree od freedom, defining the RDM $\rho_{\rm LR}^{\rm orb}=\sum_{\alpha,\beta} \la \alpha,\beta|\rho_{\rm LR}|\alpha,\beta \ra \in 4 \times 4$ for two orbital qubits.
The Fig.~\ref{fig-3} shows results for the corresponding average concurrence 
$\la C_{\rm LR}^{\rm orb} \ra$ vs $L_{\rm esc}/L_{\rm SO}$  for parallel 
($s_1=s_2=\ua$; solid line) and antiparallel ($s_1=\ua, s_2=\da$; dashed line) incoming 
spins subject to a magnetic flux $\phi$. The insets depict the fluctuations, respectively. 
Outgoing spins of different species do not contribute to orbital entanglement. 
\cite{samuelsson05} This is why only parallel incoming spins show a finite 
$\la C_{\rm LR}^{\rm orb} \ra \approx 0.39$ in the absence of SO coupling 
($L_{\rm esc}/L_{\rm SO}=0$) when leaving the quantum dot. This value, almost unaffected by 
the breaking of TRS (finite $\phi$), is in very good agreement with previous RMT 
results\cite{beenakker03-2} 
depicted by the full dots in Fig.~\ref{fig-3}. 
As SO coupling increases, spins flip during transport and outgoing spin channels of 
different sign open up both at the left and right of the dot. This hinders 
the production of orbital entanglement from originally parallel spins, leading to a 
reduction of the concurrence. In contrast, for incoming antiparallel spins SO scattering 
contributes to the formation of orbitally entangled states between spins of the same 
outgoing species. The scale on which the concurrence varies significativelly as a function 
of $L_{\rm esc}/L_{\rm SO}$ is similar to that determining the transition fron WL to WA in 
the conductance shown in Fig.~\ref{fig-2}. 
For large SO coupling, the degree of entanglement saturates 
as the orientation of outgoing spins randomize. However, 
a finite difference  $\Delta C \approx 0.05$ survives between 
different incoming spin configurations for $\phi=0$ (Fig.~\ref{fig-3}, upper panel). 
This is a consequence of
the TRS correlations preserved by the SO interaction: As soon as a finite $\phi$ breaking 
TRS is applied the concurrence tends rapidly to a common
asymptotic value independent of the initial condition (Fig.~\ref{fig-3}, lower panel). 
This indicates that breaking time-reversal and spin-rotation symmetries 
give rise to a residual orbital entanglement with universal average 
concurrence $\la C_{\rm LR}^{\rm orb} \ra \approx 0.075$ for chaotic dots. 
We point out that
orbital entanglement is very sensitive to the spin dynamics even for broken TRS  
(Fig.~\ref{fig-3}, lower panel), 
where WL and WA quantum corrections to the conductance are absent 
(dotted line in Fig.~\ref{fig-2}; see also Refs.~\onlinecite{zumbuhl02,AF01,BCH02,zaitsev05}).
Regarding the fluctuations (insets in Fig.~\ref{fig-3}), they show a functional dependence 
similar to that of the concurence. 
For large $L_{\rm esc}/L_{\rm SO}$, $\sqrt{{\rm var}(C)} \approx \la C \ra$. 
These features repeat in our results of Figs.~\ref{fig-4} and \ref{fig-5}. 
We further note that some difficulties may appear for detecting orbital 
entanglement produced from incoming antiparallel spins by violating Bell 
inequalities for shot noise as described in Ref.~\onlinecite{beenakker03-2}. This is 
because both incoming as well as outgoing channels would mix at the left side of 
the dot (unless they can be spatially separated). However, this does not exclude 
the possibility implementing alternative approaches for detection as, e.g., the 
determination of lower bounds for entanglement from beam-splitter current 
correlations\cite{BL03-GFTF06} or the use of some entanglement witness.\cite{witness}

%%%%%%%%%%%%%%%%%%%%%%%%%%%%%%%%%%%%%%%%%%%%%%%%%%%%%%%%%%%%%%%%%%%%%%%%%%%%%%%%%%%%%%%%%%
%                                       FIGURE
%%%%%%%%%%%%%%%%%%%%%%%%%%%%%%%%%%%%%%%%%%%%%%%%%%%%%%%%%%%%%%%%%%%%%%%%%%%%%%%%%%%%%%%%%%
%%%%%%%%%%%%%%%%%%%%%%%%%%%%%%%%%%%%%%%%%%%%%%%%%%%%%%%%%%%%%%%%%%%%%%%%%%%%%%%%%%%%%%%%%%

Information regarding the degree of {\it spin entanglement} between left and right 
channels contained in (\ref{outLR}) can be evaluated from $\rho_{\rm LR}$ by tracing 
out the orbital degree of freedom instead, constructing the RDM 
$\rho_{\rm LR}^{\rm spin}=\sum_{p,q} \la p,q|\rho_{\rm LR}|p,q \ra \in 4 \times 4$ 
for two spin qubits.
Results for the corresponding 
average concurrence $\la C_{\rm LR}^{\rm spin} \ra$ are presented in Fig.~\ref{fig-4}. 
In contrast to the previous case of orbital entanglement, antiparallel 
incoming spins (dashed lines) lead now to a finite 
$\la C_{\rm LR}^{\rm spin} \ra$ 
already at $L_{\rm esc}/L_{\rm SO}=0$ due to exchange correlations, 
while parallel incoming spins (solid lines) do not. 
The result is in agreement with our independent RMT simulations (full dots) 
performed by following the approach of Ref.~\onlinecite{beenakker03-2}.
The presence of multiple orbital channels give rise to an outgoing \emph{mixed} 
state ($\Tr {\rho_{\rm LR}^{\rm spin}}^2 < 1$) 
with $\la C_{\rm LR}^{\rm spin} \ra < 1$.
This differs from the case in which \emph{one} single-orbital-channel lead is 
attached at each side of the dot:
There, antiparallel incoming spins escape at the left and right in a \emph{pure} singlet 
state with $\la C_{\rm LR}^{\rm spin} \ra = 1$ independently of the scattering 
amplitudes (straightfoward from Eq.~(\ref{outLR}); see also Ref.~\onlinecite{samuelsson05}). 
We also note that $\la C_{\rm LR}^{\rm spin} \ra \ll \la C_{\rm LR}^{\rm orb} \ra$ at 
$L_{\rm esc}/L_{\rm SO}=0$ (compare Figs.~\ref{fig-3} and ~\ref{fig-4}).
This is probably related to the fact that, in contrast to spin entanglement, 
orbitally entangled electrons leave the dot at left and right in a \emph{pure} state 
($\Tr {\rho_{\rm LR}^{\rm orb}}^2 = 1$).  
For large SO coupling, Fig.~\ref{fig-4} shows features similar to those for orbital entanglement:
A finite $\Delta C$ survives between different incoming states at $\phi=0$ due to TRS correlations 
(Fig.~\ref{fig-4}, upper panel). The difference disappears as TRS is broken by a 
finite $\phi$ (Fig.~\ref{fig-4}, lower panel). More interestingly, the asypmtotic value for 
$\la C_{\rm LR}^{\rm spin} \ra$ is very similar to that for $\la C_{\rm LR}^{\rm orb} \ra$ 
in Fig.~\ref{fig-3} (lower panel). 
This indicates the existence of a universal value for the concurrence of residual 
left-right entanglement independently of the initial condition \emph{and} particular 
degree of freedom. 
The claim is supported by RMT, which shows that when both time-reversal and spin-rotation 
symmetries are broken the $S$-matrix is uniformly distributed in the unitary group and 
no effective difference exists between spin and orbital channels. \cite{beenakker97}

We consider now the entanglement production for transmitted spins, i.e., the 
spin entanglement between channels at the right side of the dot. This 
entanglement is contained in the $|0,2\ra$ component of Eq.~(\ref{sOUT}). 
We note that such component splits at the same time into other three contributions, 
each of them with different local particle number at leads 3 ($n_3$) and 
4 ($n_4$) such that $n_3+n_4=2$. Following Sec.~\ref{ParCon}, we see 
that also here the entanglement can be studied independently in each term 
due to local particle number conservation. The only component of interest 
is that one with one single excitation on each lead, obtained by projecting
the outgoing state (\ref{OUT}) onto the corresponding subspace. It reads 
\be
| \Psi_{\rm R} \ra \equiv \sum_{\alpha,\beta}
( S_{3 1}^{\alpha s_1} S_{4 2}^{\beta s_2} - S_{4 1}^{\beta s_1} S_{3 2}^{\alpha s_2} )
~b_3^{\alpha \dag}  b_4^{\beta \dag} | 0 \rangle, 
\label{outR}
\ee
where $\alpha$ and $\beta$ label ougoing spins in the leads 3 and 4, respectively.
Note that the $| \Psi_{\rm R} \ra$ can be only \emph{spin-entangled} since there are 
just two orbital channels on the right side of the dot.
After defining the density matrix 
$\rho_{\rm R}^{\rm spin}=|\Psi_{\rm R}\ra \la \Psi_{\rm R}|/\la \Psi_{\rm R}|\Psi_{\rm R}\ra$ for the two spin qubits,
we plot in Fig.~\ref{fig-5} (upper panel) the corresponding $\la C_{\rm R}^{\rm spin} \ra$ 
vs $L_{\rm esc}/L_{\rm SO}$ for $\phi=0$. 
As in the previous case of left-right spin entanglement, only antiparallel incoming 
spins (dashed line) lead to a finite $\la C_{\rm R}^{\rm spin} \ra$ 
at $L_{\rm esc}/L_{\rm SO}=0$. However, the degree of entanglement is much larger
(its value agrees with our RMT calculations depicted by the full dot).
That holds true for large SO coupling, where the concurrence arrives at a 
relatively large common asymptotic value $\la C_{\rm R}^{\rm spin} \ra \approx 0.4$ 
independently of the initial condition. 
This contrast with our findings for left-right entanglement, where TRS correlations are 
relevant.
The application of a finite $\phi$, Fig.~\ref{fig-5} (lower panel), does not 
affect the zero-flux characteristics significatively.

\begin{figure}
%%[tbp]
\includegraphics[width=.4 \textwidth, angle=0]{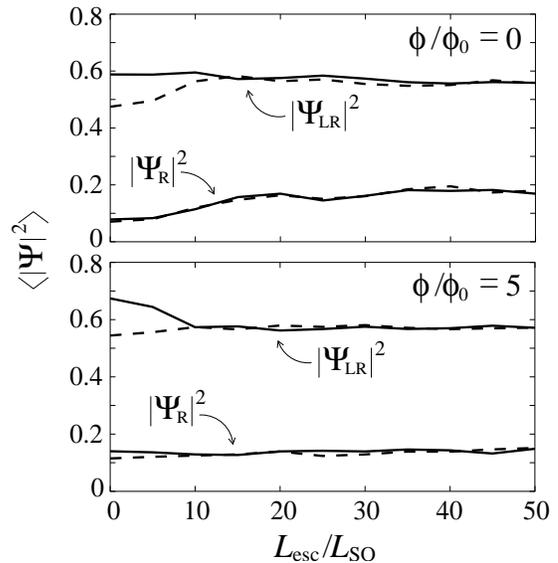}
\caption{
Sample-averaged probabilities for the states (\ref{outLR}) and (\ref{outR}) vs. 
SO coupling. 
Left (right) panel shows results for preserved (broken) TRS. Solid (dashed) lines 
correrspond to (anti)parallel incoming spins.
}
\label{fig-6}
\end{figure}

Usefull information can be still extracted from  
the wave functions (\ref{outLR}) and (\ref{outR}) by plotting their modulus square 
(Fig.~\ref{fig-6}). The larger contribution is given by $|\Psi_{\rm LR}|^2$. 
The $|\Psi_{\rm R}|^2$ is much smaller instead.
It means that highly spin-entangled electrons transmitted to the right are actually 
produced with a lower probability. This can be understood by using classical probabilities
in the limit of broken time-reversal and spin-rotation symmetries: The probability 
for the two incoming electrons to end up in any two different channels is $1/28$. 
There are 16 combinations giving one electron on the left and one on the right, and 
only 4 with one electron in each lead 3 and 4. The quantities
$16/28 \approx 0.571$ and $4/28 \approx 0.143$ are in agreement with the asymptotic 
values of Fig.~\ref{fig-6} (lower panel). Quantum corrections to these values show up for 
small SO coupling, specially when $\phi=0$. 
We further note in Fig.~\ref{fig-6} that a probability difference appears between parallel 
(solid line) and antiparallel (dashed line) incoming spins in the left-right component due 
to Fermionic statistics. Breaking TRS increases the contribution keeping the relative 
difference unaffected. 

%%%%%%%%%%%%%%%%%%%%%%%%%%%%%%%%%%%%%%%%%%%%%%%%%%%%%%%%%%%%%%%%%%%%%%%%%%%%%%%%%%%%%%%%%%
%                                       FIGURE
%%%%%%%%%%%%%%%%%%%%%%%%%%%%%%%%%%%%%%%%%%%%%%%%%%%%%%%%%%%%%%%%%%%%%%%%%%%%%%%%%%%%%%%%%%
%%%%%%%%%%%%%%%%%%%%%%%%%%%%%%%%%%%%%%%%%%%%%%%%%%%%%%%%%%%%%%%%%%%%%%%%%%%%%%%%%%%%%%%%%%

%====================================================================
\section{Conclusion}
\label{conclu}
%====================================================================

In summary, we studied the role of spin dynamics and TRS correlations 
in the production of entanglement in mesoscopic conductors
and the connection with WL and WA quantum corrections. 
By including SO coupling, among other things we found that the TRS 
effects can be more important than originally thought.\cite{beenakker03-2} 
These manifest both in the degree of entanglement as well as in the 
production rate. The effects of SO coupling appear
on a scale corresponding to the pass from the regime of WL to WA in the 
quantum conductance.
We also determined some universal characteristics of chaotic entanglers 
as the residual amount of entanglement produced after time-reversal
and spin-rotation symmetry breaking. 

%====================================================================
\acknowledgments
%====================================================================

We thank C.W.J. Beenakker, M. B\"uttiker, M.-S. Choi, V. Giovannetti, and F. Taddei for useful 
comments. This work was supported by the European Commission through the Spintronics 
Research Training Network, and by the ``Quantum Information'' research program of 
Centro di Ricerca Matematica ``Ennio de Giorgi'' of Scuola Normale Superiore.

%%%%%%%%%%%%%%%%%%%%%%%%%%%%%%%%%%%%%%%%%%%%%%%%%%%%%%%%%%%%%%%%%%%%%
% BIBLIOGRAPHY
%%%%%%%%%%%%%%%%%%%%%%%%%%%%%%%%%%%%%%%%%%%%%%%%%%%%%%%%%%%%%%%%%%%%%

%%%%%%%%%%%%%%%%%%%%%%%%%%%%%%%%%%%%%%%%%%%%%%%%%%%%%%%%%%%%%%%%%%%%%

\end{document}